# Give more data, awareness and control to individual citizens, and they will help COVID-19 containment

Mirco **Nanni*** (ISTI-CNR, Italy), Gennady **Andrienko** (IAIS-Fraunhofer, Germany and City University of London, UK), Albert-László **Barabási** (Northeastern University, USA), Chiara **Boldrini** (IIT-CNR, Italy), Francesco **Bonchi** (ISI Foundation, Italy and Eurecat, Spain), Ciro **Cattuto** (University of Torino, Italy and ISI Foundation, Italy), Francesca **Chiaromonte** (Sant'Anna School of Advanced Studies Pisa, Italy and Penn State University, USA), Giovanni **Comandé** (Sant'Anna School of Advanced Studies Pisa, Italy), Marco **Conti** (IIT-CNR, Italy), Mark **Coté** (King's College London, UK), Frank **Dignum** (Umeå University, Sweden), Virginia **Dignum** (Umeå University, Sweden), Josep **Domingo-Ferrer** (Universitat Rovira i Virgili, Catalonia), Paolo **Ferragina** (University of Pisa, Italy), Fosca **Giannotti** (ISTI-CNR, Italy), Riccardo **Guidotti** (University of Pisa, Italy), Dirk **Helbing** (ETH Zurich, Switzerland), Kimmo **Kaski** (Aalto University School of Science, Finland), Janos **Kertesz** (Central European University, Hungary), Sune **Lehmann** (Technical University of Denmark), Bruno **Lepri** (FBK, Italy), Paul **Lukowicz** (DFKI, Germany), Stan **Matwin** (Dalhousie University, Canada and Polish Academy of Sciences, Poland), David **Megías Jiménez** (Universitat Oberta de Catalunya), Anna **Monreale** (University of Pisa, Italy), Katharina **Morik** (TU Dortmund University, Germany), Nuria **Oliver** (ELLIS Alicante, Spain and Data-Pop Alliance, USA), Andrea **Passarella** (IIT-CNR, Italy), Andrea **Passerini** (Università degli Studi di Trento), Dino **Pedreschi** (University of Pisa, Italy), Alex **Pentland** (MIT, USA), Fabio **Pianesi** (EIT Digital, Italy), Francesca **Pratesi** (University of Pisa, Italy), Salvatore **Rinzivillo** (ISTI-CNR, Italy), Salvatore **Ruggieri** (University of Pisa, Italy), Arno **Siebes** (Universiteit Utrecht, The Netherlands), Vicenc **Torra** (Maynooth University, Ireland and Umeå University, Sweden), Roberto **Trasarti** (ISTI-CNR, Italy), Jeroen **van den Hoven** (TU Delft, The Netherlands), Alessandro **Vespignani** (Northeastern University, USA)

* contact author: mirco.nanni@isti.cnr.it

April 14, 2020

**[Abstract]** The rapid dynamics of COVID-19 calls for quick and effective tracking of virus transmission chains and early detection of outbreaks, especially in the "phase 2" of the pandemic, when lockdown and other restriction measures are progressively withdrawn, in order to avoid or minimize contagion resurgence. For this purpose, contact-tracing apps are being proposed for large scale adoption by many countries[1]. A centralized approach, where data sensed by the app are all sent to a nation-wide server, raises concerns about citizens' privacy and needlessly strong digital surveillance, thus alerting us to the need to minimize personal data collection and avoiding location tracking.
We advocate the conceptual advantage of a decentralized approach, where both contact and location data are collected exclusively in individual citizens' "personal data stores", to be shared separately and selectively (e.g., with a backend system, but possibly also with other citizens), voluntarily, only when the citizen has tested positive for COVID-19, and with a privacy preserving level of granularity. This approach better protects the personal sphere of citizens and affords multiple benefits: it allows for detailed information gathering for infected people in a privacy-preserving fashion; and, in turn this enables both contact tracing, and, the early detection of outbreak hotspots on more finely-granulated geographic scale. The decentralized approach is also scalable to large populations, in that only the data of positive patients need be handled at a central level. Our recommendation is two-fold. First to extend existing decentralized architectures with a light touch, in order to manage the collection of location data locally on the device, and allow the user to share spatio-temporal aggregates - if and when they want and for specific aims - with health authorities, for instance. Second, we favour a longer-term pursuit of realizing a Personal Data Store vision, giving users the opportunity to contribute to collective good in the measure they want, enhancing self-awareness, and cultivating collective efforts for rebuilding society.

---

[1] https://www.top10vpn.com/news/surveillance/covid-19-digital-rights-tracker/

**[Introduction]** National authorities are currently addressing the rapid spread of SARS-CoV-2 through strong, sometimes extreme control measures aimed at containing and slowing the diffusion of the virus to levels that can be managed by health-care and socio-political institutions. Knowledge of where and when the diffusion is taking place is essential for shortening the emergency period, and for focusing stronger countermeasures where they are actually needed. The imposition of generalized and strong emergency measures limiting citizen liberty is, in part, an effect of our poor knowledge of how and where exactly the virus is circulating and outbreaks are growing. Very recent results[2] have clearly shown that immediate tracing of infected people - ideally from the pre-symptomatic phase - could significantly contribute to reducing the individual's infection rate (the well-known $R_0$ index) below 1.

Personal big data, able to describe the movement of people in greater detail, should be seen as a potentially powerful weapon in combatting the pandemic, for example in contact tracing, that is, revealing the places a patient who has tested positive has visited in recent days (thus identifying places in risk of contagion) in addition to the people the user has been in contact with (thus identifying specific people at risk). This kind of operation potentially puts individual privacy at risk, and is at the core of current debates about the best trade-off between privacy and data value for public health. One example is South Korea, which made the movement of positive patients de facto public, clearly favouring data value (with excellent results in containment of virus spread) while sacrificing patient privacy (who risks a social stigma, potentially dissuading people from exposing and testing for the virus[3]). In contrast, various European efforts lean toward strong individual personal data protection, stipulating clear requirements that data collection apps should satisfy[4] and promoting a unified European approach[5]. We believe it is possible to reap the benefits of contact tracing, including location data with a privacy preserving level of granularity, without forgoing personal data protection altogether while enhancing trust. GDPR compliance, abiding to the principle of privacy/data protection by design and privacy/data protection by default, enables the benefits of location data.

---

[2] Ferretti, L., Wymant, C., Kendall, M., Zhao, L., Nurtay, A., Bonsall, D.G., and Fraser, C. (2020). Quantifying dynamics of SARS-CoV-2 transmission suggests that epidemic control is feasible through instantaneous digital contact tracing. *Science*.
https://science.sciencemag.org/content/early/2020/03/30/science.abb6936
[3] https://www.nature.com/articles/d41586-020-00740-y
[4] https://www.ccc.de/en/updates/2020/contact-tracing-requirements
[5] https://www.politico.eu/article/coronavirus-europe-data-regulator-calls-for-pan-european-covid-19-app/

**[Existing proposals & their limitations]** Learning from current success stories as well as controversies, various teams of researchers and developers are now proposing a different vision where privacy protection is a *must*, and solutions are designed to extract useful data without sharing personal sensitive information. In particular, the spatial information associated with the individual citizens (where they stay or move) is considered to be too sensitive, and difficult to protect. An important research direction is the privacy-safe, spatially-oblivious implementation *proximity-tracing*, that in this context basically represents the ability to reconstruct the close contacts with other people that an individual had before being tested positive. A central example is the DP3T (Decentralized Privacy-Preserving Proximity Tracing[6]) approach. In general terms, the solution is based on mobile phone apps that continuously collect the list of anonymous, app-generated IDs of other phones (which, therefore, need to have the app installed, too) that had close and prolonged contacts with the device. With DP3T, the trusted authority simply broadcasts the anonymous app-generated IDs of the positive patient's phone, and each contact needs to check the list to find themselves. DP3T is part of a broader initiative, named PEPP-PT (Pan-European Privacy-Preserving Proximity Tracing[7]), that covers various different approaches to the problem, like replacing the broadcasting phase with a different communication way where positive users provide the list of "contacted" anonymous app-generated IDs to the trusted central authority, who is then able to directly call and warn the phones in the list. The strong point of this approach lies in the simplicity of the information used, enabling an easy and rapid implementation that guarantees privacy protection. While we believe that this approach is on the right track and is particularly useful in the short term, we also emphasize that limiting the analysis to simple contact (close-range proximity) data limits the efficacy. For example, the discoverability of potentially exposed contacts is by design limited to those who have the app installed, making it impactful only after a critical mass of users is reached. Furthermore, only direct contacts are detected, thus not considering surface-touch contamination, which is a typical phenomenon in large shared spaces, like supermarkets and such, considered to be a potential vector of diffusion[8]. Spatial information, joined with its temporal dimension, is a key ingredient to detect outbreak hotspots. Spatial-temporal information within a privacy preserving architecture (e.g. appropriate granularity levels, clear access rights and aims for data processing, enhanced security, etc.) can provide vital granular aggregate data with a modest or null impact on fundamental rights and freedoms, see for example the MIT Private Kit Safe Path initiative[9].

---

[6] https://github.com/DP-3T/documents
[7] https://www.pepp-pt.org/
[8] van Doremalen, N., *et al*. (2020). Aerosol and surface stability of HCoV-19 (SARS-CoV-2) compared to SARS-CoV-1. *The New England Journal of Medicine*. DOI: 10.1056/NEJMc2004973.
[9] Raskar, R., *et al*. (2020). Apps gone rogue: Maintaining personal privacy in an epidemic. arXiv preprint arXiv:2003.08567

**[Our proposal]** Our claim is very simple: limiting data flow at its very source is not the best answer. Individual citizens - and only them - should be able to collect detailed information about their own position and movement, together with other types of data, including (in the direction of previous proposals) pseudo-IDs of devices at close distance. The means for safeguarding privacy should instead be in **providing the users with full control of such data**, together with the necessary tools for sharing only the information they want at the preferred level of detail, to customise sharing of information depending on the individuals/entities with whom they are sharing, and for evaluating pros and cons of each sharing option.

The paradigm we envision is based on a Personal Data Store[10,11,12] where users collect and manage all their own data, equipped with data management and analytics tools for elaborating them, as well as with functionalities for controlling what kind of information - raw or derived from data - should be shared with other users or with authorities. The main points of the approach we envision, based on such environment, are the following:

- Each user has a **personal software environment** (either directly on the smart phone or in the cloud) where they can store, elaborate and control their own data in an exclusive way. No third-party has access to this data.
- The personal software environment of the user is **a tool they can actively decide to use to perform actions**, for instance to help in providing correct information to health authorities in case they are tested positive to COVID-19; or simply to contribute to public safety by joining some collective computation of global statistics useful to improve countermeasures.
- When the user decides to share information, they define the **aggregates to share**, taking into consideration the minimum spatial and temporal granularity of the information needed to realize the service. The environment provides the functionalities to define the minimum data requirement and to compute and share the data. A key point is that deciding the best trade-off between privacy and data utility might require a knowledge that is available only late in the process, because the context might change either the utility of a given type of data or the priorities of the individual. Interestingly, note that recent research has found that over 75% of individuals who reported being positive for COVID-19 had been in close contact with another individual - who they knew - infected by COVID-19[13].
- The information sharing can happen in two modalities:
    - A **simple transfer** to a trusted authority of the minimum data needed to realize the service, for instance the list of close contacts (e.g. as hashed mac-addresses

---

that only the contact herself can recognize) similar to PEPP-PT and DP3T; or the list of locations visited, in the form of Points of Interest or municipalities, useful to find potential outbreaks hotspots.
    ○ When possible, through a collaborative, **distributed computation** of global aggregates involving the information of the user, e.g. using secure multi-party computation techniques[14] or specific privacy-preserving distributed methods[15].
- The information provided by authorities (possibly thanks to the individual contributions) about risky areas and possible contacts with positive patients can be joined with the complete information the user has about themselves, providing a data analytics-enabled **self-awareness** of own behaviour and the potential points of risk. For instance, the user might not realize that their own daily home-work routine involves passing through an area that has an increased level of risk, and the analysis might suggest modifying the route to work. Note that such level of detailed information would only be available locally to the specific user, as it comes from "merging" global information provided by centralised entities (e.g., a nation-wide authority) with local information available only to the individual user (on her/his devices).

This proposal (like the PEPP-PT initiative) exemplifies a distinctively European approach of aiming to co-realize important moral obligations - in this case for public health and saving lives, together with respecting rights and fundamental freedoms - instead of choosing for a quick solution that relativizes one of our conflicting obligations.

Maintaining the trust of citizens at a time of crisis like the current one is a priority. This includes respecting the requirements for maintaining fundamental respect for human rights, ethical principles and existing legislation. A user-centric approach will also ensure that data is only used during the duration of the crisis and that the user has the control to end the tracking once the need is over. Our proposal leverages both the respect for individual freedoms and for the environment, by cultivating feelings of solidarity and a sense of collective responsibility for rebuilding society.

**[Recommendations]** Summarizing, our view is that emergency situations like the COVID-19 pandemic represent a strong case - yet one not unique - where providing people complete control of the data they produce and collect and how they are shared (and, maybe, an improved awareness of what they are collecting) can provide an edge in facing complex challenges. Initiating (centralized or decentralized) data collection with rigid predefined privacy and data quality requirements and excluding the human from the decision loop can often be suboptimal.

---

[14] Lindell, Y., and Pinkas, B.. (2008). Secure Multiparty Computation for Privacy-Preserving Data Mining. IACR Cryptology ePrint Archive. 2008. 197. 10.29012/jpc.v1i1.566.
[15] Achieving Privacy-preserving Distributed Statistical Computation. PhD thesis, 2012. https://www.research.manchester.ac.uk/portal/en/theses/achieving-privacypreserving-distributed-statistical-computation(6831db5c-d605-4a38-9711-7592d2b94e01).html

Our main recommendation, therefore, is to work on two parallel tracks: in the short-term and in the long-term. In the short term, the decentralized architectures currently under development for social contact tracing (in particular, PEPP-PT/D3PT) should be extended to manage the collection of location data locally on the device. A loose integration between the two components (contact and location data) should be provided, that keeps them logically independent and mutually not linkable. This allows us to maintain all the privacy-by-design benefits of the contact tracing solutions mentioned above, but the moment users are confirmed as a positive case, it allows them to voluntarily provide additional contextual information in a privacy-preserving way (on top of independently triggering the PEPP-PT/ D3PT contact tracing mechanisms), that can contribute to the computation of useful global aggregates[16], such as spatio-temporal density maps[17] to identify potential infection hubs through location data.

Over a longer-term, deep-impact actions should be investigated to realize a Personal Data Store approach, to enable more effective emergency countermeasures based on a novel connection between collective good and the huge information treasure that each individual brings with themselves.

---

[16] https://ec.europa.eu/info/files/recommendation-apps-contact-tracing_en (e.g. point 1.2, page 8)
[17] Anna Monreale, Wendy Hui Wang, Francesca Pratesi, Salvatore Rinzivillo, Dino Pedreschi, Gennady L. Andrienko, Natalia V. Andrienko: Privacy-Preserving Distributed Movement Data Aggregation. AGILE Conf. 2013: 225-245